\def\lb{ \left[ }
\def\rb{ \right]  }

\def\bar{\overline}
\def\hat{\widehat}
\def\*{\star}
\def\[{\left[}
\def\]{\right]}
\def\({\left(}
\def\){\right)}
\def\BR{\Bigr)}  \def\BL{\Bigr(}
\def\BBL{\lb}    \def\BBR{\rb}
\def\frac#1#2{{#1 \over #2}}
\def\inv#1{{1 \over #1}}

\def\d{\partial}

\def\bra#1{{\langle #1 |  }}
\def\ket#1{ | #1 \rangle}

\def\2pi{\hbox{$2\pi i$}}

\def\dsl{\raise.15ex\hbox{/}\kern-.57em\partial}
\def\Dsl{\,\raise.15ex\hbox{/}\mkern-.13.5mu D}

\def\det{ {\rm det}}

\def\th{\theta}		
\def\ga{\gamma}

\def\ep{\epsilon}
\def\la{\lambda}	\def\La{\Lambda}
\def\de{\delta}		\def\De{\Delta}
\def\om{\omega}		\def\Om{\Omega}
\def\sig{\sigma}	

\def\CP{{\cal P}}		
	\def\CT{{\cal T}}

\font\numbers=cmss12
\font\upright=cmu10 scaled\magstep1
\def\stroke{\vrule height8pt width0.4pt depth-0.1pt}
\def\topfleck{\vrule height8pt width0.5pt depth-5.9pt}
\def\botfleck{\vrule height2pt width0.5pt depth0.1pt}
\def\Zmath{\vcenter{\hbox{\numbers\rlap{\rlap{Z}\kern
		0.8pt\topfleck}\kern
		2.2pt \rlap Z\kern 6pt\botfleck\kern 1pt}}}
\def\Qmath{\vcenter{\hbox{\upright\rlap{\rlap{Q}\kern
                   3.8pt\stroke}\phantom{Q}}}}
\def\Nmath{\vcenter{\hbox{\upright\rlap{I}\kern 1.7pt N}}}
\def\Cmath{\vcenter{\hbox{\upright\rlap{\rlap{C}\kern
                   3.8pt\stroke}\phantom{C}}}}
\def\Rmath{\vcenter{\hbox{\upright\rlap{I}\kern 1.7pt R}}}
\def\Z{\ifmmode\Zmath\else$\Zmath$\fi}
\def\Q{\ifmmode\Qmath\else$\Qmath$\fi}
\def\N{\ifmmode\Nmath\else$\Nmath$\fi}
\def\C{\ifmmode\Cmath\else$\Cmath$\fi}
\def\R{\ifmmode\Rmath\else$\Rmath$\fi}
\def\cadremath#1{\vbox{\hrule\hbox{\vrule\kern8pt\vbox{\kern8pt
			\hbox{$\displaystyle #1$}\kern8pt}
			\kern8pt\vrule}\hrule}}

\def\presentation{
\voffset -.50in   %\voffset -1.05in
\hoffset -.19in
\oddsidemargin 0in \evensidemargin 0in
\marginparwidth .75in \marginparsep 7pt \topmargin 0in
\headheight 12pt \headsep .25in
\footheight 18pt \footskip .35in
\textheight 9.5in \textwidth 6.5in
\columnsep 10pt \columnseprule 0pt }
\def\debut{ \begin{eqnarray} }
\def\fin{ \end{eqnarray} }
\def\non{ \nonumber }
\documentstyle[12pt]{article}
\presentation
\begin{document}
\rightline{SPhT-94-119}
\rightline{hep-th/9411017}
\vskip 1cm
\centerline{\LARGE Some Simple (Integrable) Models of Fractional Statistics.
\footnote[1]{Presented at Les Houches Summer School on "{\it
Fluctuating Geometries in Statistical Mechanics}", 2 Aug-9 Sep, 1994.}}
\bigskip
%\centerline{\LARGE ~~~~~~~~ titre ~~~~~~~~~~~~~~~~~~~~~~~~~}
%\vskip 1cm
%
%\centerline{\large Olivier Babelon }
%\centerline{Laboratoire de Physique Th\'eorique et Hautes
%Energies \footnote[1]{\it Laboratoire associ\'e au CNRS.}}
%\centerline{ Universit\'e Pierre et Marie Curie, Tour 16 1$^{er}$
%\'etage, 4 place Jussieu}
%\centerline{75252 Paris cedex 05-France}
%
\vskip1cm
\centerline{\large  Denis Bernard
\footnote[2]{Member of the CNRS} }
\centerline{Service de Physique Th\'eorique de Saclay
\footnote[3]{\it Laboratoire de la Direction des Sciences de la
Mati\`ere du Commisariat \`a l'Energie Atomique.}}
\centerline{F-91191, Gif-sur-Yvette, France.}
\vskip 3cm
Abstract.\\
In the first part, we introduce the notion of fractional statistics in
the sense of Haldane. We illustrate it on simple models related to anyon
physics and to integrable models solvable by the Bethe ansatz.\\
In the second part, we describe the properties of the long-range interacting
spin chains. We describe its infinite dimensional symmetry, and we explain
how the fractional statistics of its elementary excitations is an echo of
this symmetry.\\
In the third part, we review recent results on the Yangian representation
theory
which emerged from the study of the integrable long-range interacting models.
\vfill
\newpage
%
%%%% DEBUT  %%%%%%%%%%%%%%%%%%%%%%%%%
%
%
\section{Haldane's fractional statistics. }

\subsection{Definition. }

Haldane \cite{Ha1} has recently introduced a notion of fractional
statistics which is independent of the dimension of space.
This notion is not based on the monodromy properties of
the $N$-particle wave functions, but on the way the
number of available single-particle states varies when
particles are added into the system. More precisely,
consider a system with a total number of particles $N=\sum_jN_j$,
with $N_j$ the number of particles of the species $j$.
Consider now adding a particle of the species $i$ into
the system without changing its size and the boundary conditions.
Keeping fixed the positions of the $N$ particles of the
original system, the wave function of the new $(N+1)$-body
system can be expanded in a basis of wave functions for the added
particle. We denote by $D_i$ the dimension of this basis.
The important point is that this dimension may depend
on the numbers $N_j$ of particles in the original system.
Assuming that this dependence is linear,
Haldane defines the ``{\it statistical interaction}"
through the relation
\cite{Ha1}~:
\debut
\frac{\d D_i}{\d N_j} = - g_{ij}~. \label{eAa}
\fin
Clearly, for bosons the numbers of available single-particle states
are independent of the numbers $N_j$, and $g_{ij}=0$.
For fermions, the numbers of available single-particle states
decrease by one for each particle added, and $g_{ij}=\delta_{ij}$.

One of the ideas underlying the introduction of the generalized
Pauli principle (\ref{eAa}) is the fact that bosons and fermions can
be considered on an equal footing as far as state counting is
concerned. Indeed, for bosons or fermions, the number of states
of $N$ identical particles distributed among $G$ accessible
orbitals can be written in a unified way as~:
\debut
W_{b,f}=\frac{ (D_{b,f}+N-1)!}{N!\,(D_{b,f}-1)!} \non
\fin
with $D_b(N)=G$ for bosons, and $D_f(N)=G-N+1$ for fermions.
The dimensions $D_{b,f}$ are the numbers of accesible
states for the $N^{th}$ particle to be added. As it should be, $D_f$
decreases by one unit each time a fermion is added.
This is generalized to fractional statistics by assuming that the total
number of states with $\{N_j\}$ particles is~:
\debut
W = \prod_i \frac{[D_i(\{N_j\})+N_i-1]!}
{N_i!\,[D_i(\{N_j\})-1]!}~,
\label{EAb}
\fin
where $D_i(\{N_j\})$ is obtained by integrating (\ref{eAa})~:
\debut
D_i(\{N_j\})+\sum_j g_{ij}\, N_j =G^0_i~, \label{EAc}
\fin
with $G^0_i\equiv D_i(\{0\})$ a constant, which is interpreted as
the number of available single-particle states when
no particle is present in the system. Namely, $G^0_i$
are the bare numbers of single-particle states.

\subsection{Anyon-inspired examples.}

The first example \cite{Ha1} is a very naive example. Let us model anyons
as charged particles carrying a magnetic flux. Consider now
a collection of $\{N_i\}$ anyons with charges $q_i$ and flux $\phi_i$ on
a disc through which goes a total magnetic flux $\phi_B$.
If we assume that the charged particles do not interact, the
number $D_i$ of accessible states for a particle of charge $q_i$ in
the disc is~:
\debut
D_i \simeq q_i \frac{\phi_B}{\phi_0} \non
\fin
where $\phi_0=h/e$ is the flux quantum. Since
the anyons carry magnetic flux, each time we introduce an anyon in the system
we increase the total magnetic flux by an amount equal to
the anyon flux. Therefore, the total magnetic flux is linear in
the anyon numbers~: $\phi_B \simeq \phi_B^0 + \sum_j \phi_j N_j$.
As a consequence, the numbers of accessible states depend on the number
of anyons present in the system, and we have~:
\debut
g_{ij} = -\frac{\d D_i}{\d N_j} = -\frac{q_i\phi_j}{\phi_0} \non
\fin
The quantity $\th_{ij}=\pi(g_{ij}+g_{ji})$ coincides with the Bohm-Aharonov
phase obtained by moving an anyon of charge $q_i$ and flux $\phi_i$
around another anyon of charge $q_j$ and flux $\phi_j$.

The second example is based on a study of the elementary excitations
of the fractional quantum Hall effect \cite{Exp}.
So, we consider electrons moving on a plane in which there is
a uniform transverse magnetic field $B$. We denote by $z=x+iy$ the complex
coordinate in this plane, and by $l_B=\sqrt{h/\pi eB}$ the magnetic length.
If the electrons are not interacting, the energy spectrum is described by
the Landau levels. In an appropriate gauge, a basis of wave functions
in the first Landau level is given by~:
$\psi_j(z,\bar z) = z^j~\exp(-|z|^2/2l_B^2)$.
These functions are peacked around the circle $|z|^2=j~l^2_B$.
To confine the electrons inside a disc of radius $R$, we impose
that $j~l^2_B \leq \pi R^2$. This is equivalent to imposing
that $j \leq N_\phi$ where $N_\phi$ is the
number of flux quanta going through the disc, i.e.
$\phi_B=\pi R^2B=N_\phi\phi_0$.
If there are $N_e$ electrons in the system, the filling factor $\nu$
is defined as~:
\debut
\nu = \frac{N_e}{N_\phi} =
\frac{ \sharp ~~ of~~ electrons}{\sharp ~~ of~~ flux~~ quanta} \label{filling}
\fin

If the electrons are subject to the coulomb interaction, the
degeneracy between the Landau states is removed.
The energy spectrum was numerically studied in ref.\cite{Numer}.
It can be described as follows.  Suppose that the
filling factor $\nu$ is  slightly less than the fraction $1/m$,
namely:
\debut
N_\phi = m(N_e-1) + n \label{nufrac}
\fin
for some integer $n$.
Then, there will be a gap in the energy spectrum in the first Landau level.
The gap is of order $e^2/l_B$. There are a large number of states
above the gap but only few below the gap. These low energy states
are the accessible states for $n$ excitations carrying each a
unit flux quantum. These excitations are called quasi-holes.
The number of accessible states for the quasi-hole excitations
can be understood using Laughlin anstaz for the quasi-hole wave functions
\cite{Lau}, which for a filling factor just below $1/m$ as in eq.(\ref{nufrac})
are defined by~:
\debut
\Psi(z)= e^{-\sum_j|z_j|^2/2l_B^2}~ \prod_{i<j}(z_i-z_j)^m~ P(z) \label{wave}
\fin
where $z_j$ are the coordinates of the $N_e$ electrons, and $P(z)$ is a
polynomial, symmetric in these variables. These wave functions are
ansatz for the states which are below the gap.
So different choices of polynomial corresponds to different accessible
states for the quasi-hole excitations. These polynomials are constrained
by the fact that the electrons are confined in the disc; i.e. the
electron angular momenta should be less than $N_\phi$. Therefore,
\debut
{\rm degree}(P) + m(N_e-1) \leq N_\phi \non
\fin
or equivalently, the  degree of $P$ is less than the number of quasi-holes.
Hence, the number of accessible states to the quasi-holes is the number
of symmetric polynomials in $N_e$ variables and of degree less than
$n$ in each of the variables. This number is $\frac{(N_e+n)!}{N_e!n!}$.
It agrees with the numerical simulations \cite{Numer}.
Comparing with eq.(\ref{EAb}), we see that, at fixed number of
flux quanta $N_\phi$, the effective dimension of the quasi-hole Hilbert space
is:
\debut
D_{qh}(n) = 2 + \inv{m}(N_\phi - n) \label{effdim}
\fin
It depends linearly in the quasi-hole number. Eq.(\ref{effdim}) gives
the quasi-hole statistical interaction~:
\debut
g_{qh}=-\frac{\d D_{qh}}{\d n} = \inv{m} \label{qhfrac}
\fin

These examples illustrate how low energy collective excitations
may possess a statistics with a fractional character.

\subsection{The Bethe ansatz and fractional statistics.}

We now give another example based on integrable models \cite{WuBe}. We will
describe how the Bethe ansatz equations can be reinterpreted in
such way that they code a statistical interaction between
particles, assuming that particles of different momenta belong
to different species. Although it is useful in revealing
the fractional statistics of the elementary excitations of the
exactly solvable models, this remark
does not provide a new way of solving the Bethe ansatz equations.

In the Bethe ansatz approach to integrable models, the information
is encoded in the two-body $S$-matrix. We denote it by $S(k)$
with $k$ the relative momentum of the scattering particles;
we have $S(k)=- \exp(-i\th(k))$, where $\th(k)$ is the phase
shift and it is odd in $k$. The eigenstates of the $N$-body
hamiltonian (with periodic boundary condition) are labeled
by N (pseudo-)momenta $\{k_r\}$ $(r=1,,\cdots,N)$, which are solutions of
the Bethe ansatz equations~:
\begin{eqnarray}
e^{ik_rL} = \prod_{s\not= r} S(k_s-k_r)\quad
{\rm for\ all}\ r~,
\label{EBa}
\end{eqnarray}
with $L$ the length of the system. The energy $E$ of the eigenstate
$\ket{k_1,\cdots,k_N}$ is $E=\sum_r \ep^0(k_r)$,
where $\ep^0(k)$ is some universal function, e.g. $\ep^0(k)=k^2$.

The information about the statistical interactions is encoded in the
Bethe ansatz equations (\ref{EBa}), if it is rewritten
in appropriate form. Indeed, taking as usual the
logarithm of eq.(\ref{EBa}),  the Bethe ansatz
equations become~:
\begin{eqnarray}
L\,k_r= \sum_{s\not= r} \th(k_r-k_s) + 2\pi\, I_r~,
\label{EBf}
\end{eqnarray}
with $\th(k)$ the phase shift. Here $\{I_r\}$ $(r=1,\cdots,N)$
is a set of integers or half integers, depending on
N being odd or even, which one may choose to serve
as the quantum numbers labeling the eigenstates, instead of
the momenta $\{k_r\}$. For simplicity, let us consider the
cases when these $I_r$'s have be chosen to be all different,
(this is the usual situation).
The ground state corresponds to an equidistribution of
the integral quantum numbers $\{I_r\}$ in an
interval centered around the origin~: $I_{r+1}-I_r=1$. The excited
states correspond to particle/hole excitations
in the integral lattice for the $\{I_r\}$:
\begin{eqnarray}
I_{r+1}-I_r=1+M_r^h~, \label{eBf}
\end{eqnarray}
where $M_r^h$ is the number of holes, i.e. unoccupied
integer numbers, between $I_{r+1}$ and $I_r$.
The description of the states by the quantum numbers $\{I_r\}$
is a fermionic description since there cannot be two integers
taking the same value. In this decription, the statistics is simple,
but the energy is a complicated function of the  $\{I_r\}$.

We now change to the momentum description in the thermodynamical limit,
$N\to\infty$ at fixed density $D=N/L$.
It is convenient to introduce the variable $x=r/N$ which varies
from zero to one. The pseudo-momenta $k_r$, the integers $I_r$ and the numbers
of holes $M_r$ are all functions of $x$. As usual, we define the density
$\rho(k)$ of
particles of pseudo-momentum $k$ and the density of holes $\rho_h(k)$ by~:
\debut
\rho(k)&=&\inv{L(k_{r+1}-k_r)}= D\({\frac{dx}{dk}}\) \non\\
\rho_h(k)&=&\frac{I_{r+1}-I_r-1}{L(k_{r+1}-k_r)}=\rho(k)M(k)\non
\fin
The density of particles is normalized by $\int dk\rho(k)=D$.
The energy is then given by~:
\begin{eqnarray}
\frac{E}{L} = \int_{-\infty}^{+\infty} dk\
\rho(k)\,\ep^0(k)~. \label{enerthermo}
\end{eqnarray}
Thus there is no interaction energy between particles of
different pseudo-momenta. Contrary to the fermion description,
in the momentum description the dynamics looks simple but
as we will see, the statistics is non-trivial.

The statistical interaction is hidden in the Bethe anstaz equations.
Taking the difference of eqs.(\ref{EBf}) for $r$ and $(r+1)$ in
the limit $N\to\infty$ gives \cite{YY1}~:
\begin{eqnarray}
\rho_h(k)+\rho(k) + \int_{-\infty}^{+\infty}\frac{dk'}{2\pi}
\,\phi(k,k')\,\rho(k') = \rho^0(k)=\inv{2\pi}
\label{SCy}
\end{eqnarray}
with $\phi(k,k')$ the derivative of the phase shift:
$\phi(k,k')=\th'(k-k')$.
Eq.(\ref{SCy}) is the well known thermodynamical limit of the Bethe
ansatz equations. A close comparison with the state-counting equation
(\ref{EAc}) shows that they are identical provided that the
following identification are made~:
\def\toto{ \longleftrightarrow }
\begin{eqnarray}
G^0_i/L &\toto & \rho^0(k)\equiv\inv{2\pi} \label{eBd}\\
N_i/L &\toto& \rho(k) \label{EBk}\\
D_i(\{N_j\})/L &\toto& \rho_h(k)~~. \label{EBl}
\end{eqnarray}
with the discrete sum replaced by the integral
over momentum. This last idenfication is quite natural: the hole density
$\rho_h(k)$ clearly represents the density of available
states for an additional particle to be added.
Furthermore, from eq.(\ref{SCy}) we derive the formula
of the statistical interaction  in the momentum description~:
\begin{eqnarray}
g(k, k')= \de (k-k')+ {1\over 2\pi}\,\ \phi(k,k')~.
\label{FSe}
\end{eqnarray}
This shows that the dynamical interaction, which
is summarized in the two-body phase shift, is
transmuted into a statistical interaction. Similarly as the boson-fermion
equivalence in one-dimension, Eq.(\ref{FSe}) provides an illustration
of a phenomenon familiar in one-dimension: the translation of a dynamical
equation
into a statistical equation.

A simple example is provided by the Calogero-Sutherland model \cite{CS1}.
This is a model of particles interacting through a $1/r^2$ potential.
It is integrable, and its two-body $S$-matrix is
$S(k) = -\exp\[{-i\pi(\la-1)\,sign(k)}\]$, with $\la$ the coupling constant.
Therefore, $\phi(k,k')=2\pi(\la-1)\delta(k-k')$, and the
statistical interaction is~:
\begin{eqnarray}
g(k,k')= \la \, \delta(k-k')~. \label{EBy}
\end{eqnarray}
The Bethe ansatz equation then reads~:
\begin{eqnarray}
\inv{2\pi} = \rho_h(k) + \la\rho(k)~. \label{EBu}
\end{eqnarray}
The bare energy is $\ep^0(k)=k^2$.
The coupling constant $\la$ governs the statistical
interaction~: if the density of particles of momentum $k$
increases by a unit, then the holes density decreases by $\la$.
Eq.(\ref{EBy}) shows that the statistical interaction is
purely between particles with identical momenta. In this
respect, the Calogero-Sutherland system appears clearly as
an ideal gas of particles with a fractional
statistics. This property of the Calogero-Sutherland
model is also apparent in its recently computed
correlation functions \cite{LPS}.  Finally, from eq.(\ref{EBu})
we see that the duality $\la\toto 1/\la$ \cite{Dual},
which exchanges the system with coupling
$\la$ and $1/\la$, corresponds to exchanging particles and holes.

\subsection{Thermodynamics.}

Knowing how to enumerate states, it is then possible to study the
thermodynamics. In the thermodynamic limit, the numbers of
particles $\{N_j\}$, as well as the bare numbers of available
states $\{G^0_j\}$, become infinite. But the occupation
numbers $n_i=\({N_i/G^0_i}\)$ remain finite.
The entropy is $S=k_B\log W$ with $k_B$ the Boltzman constant.
A notion of generalized ideal gas was introduced in ref.\cite{Wu1}.
By definition, a system is called a generalized ideal gas
if  (a) its total energy with $\{N_j\}$ particles is simply given by~
\debut
E=\sum_j\ N_j\, \ep^0_j \label{EAd}
\fin
with constant $\ep^0_j$, and if (b) its states are counted according to
eq.(\ref{EAb}).

For such gases, the thermodynamic potential $\Om\equiv - PV$
at equilibrium can be evaluated by minimizing
$\Om = E - TS - \sum_j \, N_j\mu_j$
with respect to the variation of the densities $n_i$.
Here $T$ is the temperature and $\mu_i$ the chemical
potential for the species $i$.
The resulting thermodynamics was described in ref.\cite{Wu1}~:
\debut
\Om = -k_BT\sum_i\, G^0_i\,\log\({\frac{1+w_i}{w_i}}\)~,
\label{EAf}
\fin
where the functions $w_i$ are determined by the equations~:
\debut
\log\({1+w_i}\) +\sum_j g_{ji}\,\log\({\frac{w_j}{1+w_j}}\)
= \frac{\ep^0_i-\mu_i}{k_BT}~.
\label{EAg}
\fin
The quantities $w_i$ possess a clear physical meaning: they are
equal to the mean value of accessible states per particles of
species $i$, $w_i=D_i(\{\bar N_j\})/\bar N_i$.

These relations completely specify the thermodynamics of the
generalized ideal gas. The other thermodynamical quantities
can be derived from the relation~:
\debut
d\Om = -SdT-\sum_iN_id\mu_i-PdV~. \non
\fin
In particular, the occupation numbers $n_i$ are obtained
from $n_iG^0_i=-\d\Om/\d\mu_i$. It gives~:
$n_i = \sum_j \(B^{-1}\)_{ij}$,
where $B$ is a matrix with entries~:
$B_{ij}= w_i\delta_{ij}+h_{ij}$, with $G^0_ih_{ij}=g_{ij}G^0_j$.

The simplest example considered in \cite{Wu1} is the ideal gas
of particles with a diagonal statistical interaction; i.e. $g_{ij}
=g\delta_{ij}$, $\mu_i=\mu$. In this case, the statistical
distribution of $n_i$ is then given by $n_i=1/(w_i+g)$, with
$w_i$ satisfying eq.(\ref{EAg})~ which now becomes:
\debut
w_i\, ^g(1+w_i)^{1-g} = \exp\((\ep^0_i-\mu)/k_BT\)~.
\label{AAd}
\fin
For $g=0$ (or $g=1$), we recover the bosonic (or fermionic)
occupation numbers. Eq.(\ref{AAd}) possesses a $g\leftrightarrow 1/g$
duality~: $w_i(T;g) = w_i(-T/g;1/g)$.
This is the analogue of the particle/hole duality of the
Calogero models which we mentioned in the previous section.

For $g\neq 0,1$, a special example with
only one energy level, is that of anyons in the lowest
Landau level. In ref.\cite{Ouvry}, the anyon
thermodynamical potential was computed in the strong magnetic field
limit using a diagrammatic expansion; in ref.\cite{Wu1},
it was derived from a state counting approach
similar to those described in section (1.2).

The Bethe ansatz solvable models also provide examples
of generalized ideal gas, since
in the momentum description the energy is given by eq.(\ref{enerthermo}),
(i.e. there is no interaction between the particles of different
momenta), and the entropy is $S=k_B\log W$ with $W$ given by
eq.(\ref{EAc}) with the correspondence (\ref{EBk},\ref{EBl}).
The thermodynamics of such
gas is called the thermodynamic Bethe ansatz (TBA).
It was developed by Yang and Yang in ref.\cite{YY1}.
The equations governing the TBA are eqs.(\ref{EAf},\ref{EAg})
with the correspondence (\ref{eBd},\ref{EBk},\ref{EBl}).
%This discussion suggests a possible genralization of
%the Fermi liquid theory by incorporating statistical
%as well as dynamical interactions. At thermodynamic
%equilibrium, the system will be described
%by the particle density $\rho(\vec k)$ and the hole density
%$\rho_h(\vec k)$. In $D$ space dimensions, the entropy
%will be assumed to have an expression similar of those of the TBA~:
%\begin{eqnarray}
%\frac{S}{L}=\int d^D\vec k
%\[{(\rho+\rho_h)\log(\rho+\rho_h)-
%\rho\log\rho- \rho_h\log\rho_h}\]~, \label{entrobis}
%\end{eqnarray}
%where the densities $\rho(\vec k)$ and $\rho_h(\vec k)$
%are assumed to be coupled with a phenomenological
%statistical interaction $g(\vec k,\vec q)$:
%\debut
%\rho_h(\vec k) + \int d^D\vec q\  g(\vec k,\vec q) \rho(\vec q)
%= \rho^0~. \label{estati}
%\fin
%Mimicing the Fermi liquid theory,
%the simplest ansatz for the energy is quadratic in the particle density~:
%\debut
%\frac{\d E}{\d \rho(\vec k)} =  \ep^0(\vec k)
%+ \int d^D\vec q\  V(\vec k,\vec q) \rho(\vec q)
%\equiv \ep_{dr}(\vec k;\rho)~, \label{efermi}
%\fin
%where $V(\vec k,\vec q)$ is a phenomenological dynamical interaction.
%These three equations completly specify the thermodynamics.
%It is probably worth trying to determine how the low temperature behavior
%of such systems dependents on the statistical interaction $g(\vec k,\vec q)$.

\def\eqn#1{ \debut #1 \fin }
\section{Long-Range Interacting Models.}

We now present another model whose elementary excitations obey
a fractional statistics.
It is the XXX spin chain with long range interaction
introduced by Haldane and Shastry \cite{chain}, see also \cite{STat}.
This is a variant of the spin half Heisenberg chain, with exchange inversely
proportional to the square distance between the spins.
It possesses the remarkable properties that its spectrum is
highly degenerate and additive, and that the elementary excitations are spin
half objects obeying a half-fractional statistics
intermediate between bosons and fermions.

There is a large family of integrable long range interacting spin chains
which are defined as follows. We consider a spin chain with
$N$ sites, labeled by integers $i,j,\cdots$ ranging from $1$ to $N$.
On each sites there is a spin variable $\sig_i$ which takes two values:
$\sigma_i= \pm$. The hamiltonians, which are all su(2) invariant,
are of the following form~:
\eqn{ H\ =\ \sum_{i\not= j}\  h_{ij} \({P_{ij}-1 }\)
\label{EAa} }
where $P_{ij}$ is the operator which exchanges the spins
at the sites $i$ and $j$. For translation invariance $h_{ij}=h(i-j)$.
Demanding the integrability of the model selects the functions $h$.
The possible choices are~:
\debut
h(x)=\cases{ \frac{\ga^2}{(\sinh\ga x)^2}, & hyperbolic~ model~($\ga$~real)\cr
 ~& ~\cr \frac{(\pi/N)^2}{\(\sin\frac{\pi x}{N}\)^2 }, &
trigonometric~model\cr ~ & ~\cr
 \CP(x), & elliptic~model.\cr} \non
\fin
where $\CP(x)$ is the Weierstrass function. When $\ga\to\infty$, the
hyperbolic model reduces to the Heisenberg spin chain:
$h_{ij}=\de_{i,j+1}+\de_{j.i+1}$, and for $\ga\to 0$, the
interaction becomes the $1/x^2$ exchange.
The hyperbolic model has not been completly solved for general $\ga$,
although a partial list of eigenstates is known.
The elliptic model is even more intriguing  since it
interpolates between the Heisenberg spin chain of finite length
and the trigonometric model \cite{Inno}.

The Haldane-Shastry spin chain is the trigonometric model.
In the thermodynamical limit, $N\to\infty$, it reduces to the
$1/x^2$ exchange model, but it also possesses remarkable properties
at finite $N$. Notably, its hamiltonian commutes with an infinite
dimensional algebra whose two first generators are \cite{nous}~:
\debut
\vec Q_0 &=& \sum_i\ \vec S_i \label{qzero}\\
\vec Q_1&=& \sum_{i\not= j}\ cotg\({\frac{\pi(i-j)}{N}}\)\
\vec S_i\times \vec S_j \label{qun}
\fin
with $\vec S_i$ the spin operators acting on the site $i$.
The first generators are the usual su(2) generators. Together with
the second ones, they form a representation of the su(2) Yangian,
(which is a deformation of the su(2) current algebra,
see section 3 for an introduction to the Yangians).
This infinite dimensional symmetry is at the origin of the large degeneracy of
the spectrum. The fact that the hamiltonian is Yangian invariant
at finite $N$ is particular to the Haldane-Shastry spin chain;
in the Heisenberg spin chain, the Yangian symmetry only appears
in the thermodynamical limit.

In order to grasp the rules describing the spectrum, we
first construct few eigenstates.
Clearly, the ferromagnetic vacuum $\ket{\Om}=\ket{++\cdots++}$
is an eigenstate~: its energy is zero.
The eigenstates in the one-magnon sector
are  the plane waves~:$\ket{k}=\sum_n\exp(i2\pi kn/N)\sig_n^-\ket{\Om}$,
with pseudo-momentun $k$, $1\leq k\leq(N-1)$:
the one-magnon energy is $\ep(k)=\({\frac{\pi}{N}}\)^2k(k-N)$.
In the two-magnon sectors, i.e. for states of the
form $\ket{\psi}=\sum_{n,m} \psi_{n,m} \sig_n^-\sig_m^-\ket{\Om}$,
the eigenstates which are not degenerate with the zero or one-magnon
eigenstates are labeled by two pseudo-momenta $k_1,k_2$,
with $1\leq k_1,k_2 \leq (N-1)$. They are given by~:
\debut
\psi^{[k_1,k_2]}_{n,m} =
(k_1-k_2)\({\om^{nk_1+mk_2}+\om^{mk_1+nk_2}}\)-
\frac{\om^n+\om^m}{\om^n-\om^m}\({\om^{nk_1+mk_2}-\om^{mk_1+nk_2}}\)\non
\fin
with $\om=\exp(i2\pi/N)$.
Note that these  wave functions vanish if
$k_1=k_2$ but also if $|k_1-k_2|=1$. The energy of $\ket{\psi^{[k_1,k_2]}}$
is $E=\ep(k_1)+\ep(k_2)$.

 From the two-magnon computation we learn two properties of
the spectrum~: (i) it is additive, e.g. the two-magnon energy is
the sum of the one-magnon energies, but (ii) the pseudo-momenta
satisfy a selection rule~: they are neither equal nor they differ
by a unit. These rules are the general rules,
and the full spectrum  can be described as follows \cite{Ha91}.
To each eigenstate multiplet is associated a set of
pseudo-momenta $\{k_p\}$ which are non-consecutive integers
ranging from $1$ to $(N-1)$.
The energy of an eigenstate $\ket{\{k_p\}}$ with pseudo-momenta $\{k_p\}$ is:
\eqn{
H \ket{\{k_p\}}\ =\ \({\sum_p \ep(k_p)}\) \ket{\{k_p\}}
\qquad {\rm with}\quad \ep(k)= \({\frac{\pi}{N}}\)^2 k(k-N)
\label{EEAb} }
Furthermore, the degeneracy of the multiplet with pseudo-momenta $\{k_p\}$
is described by its su(2) representation content as follows.
Encode the pseudo-momenta in a sequence of $(N-1)$ labels $0$ or $1$
in which the $1$'s indicate the positions of the pseudo-momenta;
add two $0$'s at both extremities of the sequence which now has
length $(N+1)$. Since the pseudo-momenta are neither equal nor
consecutive, two labels $1$ cannot be adjacent.
The sequence corresponding to the ferromagnetic vacuum is a line of $0$,
those of the one-magnon states have $N$ label $0$ and only one label $1$,
and so on.
A sequence can be decomposed into the product of elementary
motifs, which are series of $(Q+1)$ consecutives $0$'s.
The multiplicity of the spectrum is recovered if to each elementray
motif of length $(Q+1)$ we associate a
 spin $Q/2$ representation of su(2). The representation
content of the full sequence is then given by the tensor product
of its elementary motifs.

The magnons are the excitations over the ferromagnetic vacuum;
the excitations over the antiferromagnetic vacuum are conveniently
described in terms of spinons.
For $N$ even, the antiferromagnetic vacuum corresponds to
the alternating sequence of symbols $010101\cdots010$.
The excitations are obtained by flipping and moving
the symbols $0$ and $1$. Let us give the sequences corresponding
to the first few excitations over the antiferromagnetic vacuum,
(for concreteness we choose $N=10$)~:
\debut
 ~\cases{ ~0~1~0~1~0~1~0~1~0~1~0~, & antiferromagnetic vacuum (o)\cr ~&~\cr
 ~0~1~0~1~0~1~0_x0_x0~1~0~, &  a two-spinon excitations (2a)\cr ~&~\cr
 ~0~1~0_x0~1~0_x0~1~0~1~0~, & a two-spinon excitations (2b), etc...\cr } \non
\fin
We have inserted a $x$ between any two consecutive labels $0$. These
crosses represent the spinon excitations, their number is the spinon number.
Note that there is no one-spinon excitation for $N$ even.
By convention, we will say that consecutive crosses not separated by any label
$1$
correspond to spinons in the same orbital, while crosses
separated by labels $1$ correspond to spinons in different
orbitals. From the rules described above, it follows that
the degeneracy of the excitations (2a) and (2b) are different~:
it is three in the case (2a) and four in the case (2b).
These degeneracy are recovered by giving a su(2) spin half to the
spinons and by assuming that spinons in the same orbital are in a fully
symmetric states. Hence, in the case (2a), there are two spinons
in the same orbital and therefore they form a spin one representation
of su(2), and in the case (2b), the two spinons are in two different orbitals
and therefore they form a su(2) representation isomorphic to the
tensor product of two spin half representations of su(2).
The fact that the spinons are spin half excitations can also be seen
by looking at the excitations of a spin chain of length $N$ with $N$ odd.

This description of the states generalizes to the full spectrum.
We can classify the sequences by their number
$M$ of pseudo-momenta. The spinon number $N_{sp}$ of a sequence is then defined
by
$M=\frac{N-N_{sp}}{2}$.
Since $M$ is an integer, $(N-N_{sp})$ is always even~: this means
that the spinons are always created by pairs.
A sequence of pseudo-momenta $\{k_p;p=1,\cdots,M\}$,
in the $N_{sp}$ spinon sector, can be decomposed into  $(M+1)$ elementary
motifs where, as before an elementary motif is a series of consecutive $0$.
We call the elementary motifs the accessible orbitals  to
the spinons. At fixed $N_{sp}$, there are $N_{orb}=(1+\frac{N-N_{sp}}{2})$
orbitals.
Hence, a sequence of pseudo-momenta $\{k_p\}$ corresponds
to the filling of the $N_{orb}$ orbitals with
respective spinon occupation numbers $n_p=(k_{p+1}-k_p-2)$,
with $k_0=-1$ and $k_{M+1}=N+1$ by convention.
Since an elementary motif of length $(Q+1)$ corresponds
to a spin $Q/2$ representation of su(2),
the full degeneracy of the sequences is then recovered by
assuming that the spinons are spin half objects which
behave as bosons in each orbitals.

The spinons are not bosons but ``semions":  they obey a
half fractional statistics.
This follows from the fact that the number of
available orbitals varies with the total occupation number \cite{Ha1}.
Indeed, at spinon number $N_{sp}$, the number of orbitals is
 $N_{orb}=(1+\frac{N-N_{sp}}{2})$. Therefore, we have the statistical
interaction~:
\debut
g_{sp}=-\frac{\d N_{orb}}{\d N_{sp}} = 1/2 \non
\fin
The fractional statistics of the spinons is also apparent in the spin-spin
correlation function \cite{Haspin}.
In the following section, we will describe how the fractional statistics
of the spinons is encoded in the Yangian representation theory.

The spinon description of spectrum is very similar to the description
of the excitations of the Heisenberg spin chain given by Faddeev and
Takhtajan \cite{FadTak}.

Note that the model is gapless. Its low energy properties belong
to the same universality class as the Heisenberg model.
The low energy, low temperature, behavior is described
by the level one su(2) WZW conformal field
theory. The spinon formulation of the Haldane-Shastry spin chain provides
a new quasi-particle description of the states in the WZW model \cite{YangWZW}.

\section{Algebraic Solution of the Long-Range Interacting Models.}

In this section we review few of the new results on integrable
models and on the Yangian representation theory which emerged
from the study of the long-range interacting models.
But we first need to recall standard result concerning the
algebraic Bethe ansatz, cf e.g. \cite{Fa82}.

\subsection{Algebraic Bethe ansatz and Yangians.}

We introduce the basic notion of the algebraic Bethe ansatz,
using the quantum Heisenberg chain as an example.
We consider a chain of length $N$: on each site there
is a spin variable $\sig_j$. We denote by $ S^{ab}_j$, $a,b=1,~2$,
the spin operators satisfying the su(2) commutation relations~:
\debut
\BBL\ S^{ab}_j\ ,\ S^{cd}_k\ \BBR = \de_{jk}
\BL\ \de^{cb}\ S^{ad}_j - \de^{ad}\ S^{cb}_j\ \BR \label{CIIi}
\fin
The Heisenberg hamiltonian is~:
\debut
H = \sum_{k=1}^{N}\sum_{ab} S^{ab}_k S^{ba}_{k+1}
= \sum_{k=1}^{N}\( P_{k,k+1}-1\)
\label{CIIii}
\fin
Here, we have assumed periodic boundary conditions.
As is well known, in order to preserve the integrability
the  spin operators $S^{ab}_k$ should act on the spin half
representation of su(2). So, the spin variables take only
two values, $\sig_j=\pm$, and
the operator $S^{ab}_j$  which acts only the $j^{th}$ spin
is represented by the canonical matrix $\ket{a}\bra{b}$.

The algebraic Bethe ansatz goes in few steps.

$\bullet)$ The first step consists in constructing the
local monodromy matrices $T_j(u)$. These matrices are $2\times 2$
matrices whose elements $T_j^{ab}(u)$ are operators.
The matrices $T_j(u)$ are defined by~:
\debut
T^{ab}_j(u)= u \de^{ab} +\la S_j^{ab} \label{local}
\fin
where $u$ is a complex number, called the spectral parameter,
and $\la$ a coupling constant. Note that the matrix $T_j(u)$
only acts on the $j^{th}$ spin. The important point is that
we can compute the commutation relations between its matrix elements.
These relations can be gathered into the famous relations of the
algebraic Bethe anstaz, see e.g. \cite{Fa82}:
\debut
R(u-v) (T(u)\otimes 1) (1\otimes T(v))
=( 1\otimes T(v))(T(u)\otimes 1) R(u-v)
\label{CIIiv}
\fin
where $R(u)$ is Yang's solution of the Yang-Baxter equation,
$R(u) = u-\la\ P$,
with $P$ the exchange operator $P(x\otimes y)= y \otimes x$.

$\bullet \bullet$) The second step consists in constructing
the complete monodromy matrix, which we denote by $T(u)$.
It is obtained by taking
the ordered product of the local monodromy matrices. Namely,
\debut
T^{ab}(u)
 &=&  \sum_{a_2\cdots a_N}\
T_1^{a a_2}(u)~T_2^{a_2 a_3}(u)~\cdots~ T_N^{a_N b}(u)
\label{CIIIiii}
\fin
It admits an $(\inv{u})$-expansion:
\debut
u^{-N}~T^{ab}(u) = \delta^{ab} + \frac{\lambda}{u}\(\sum_k S^{ab}_k\)
 +\frac{\lambda^2}{u^2}\( \sum_{j<k} \sum_d S^{ad}_j S^{db}_k\) +\cdots
\nonumber
\fin
The crucial point is the fact that the complete monodromy
matrix (\ref{CIIIiii}) satisfy the relations (\ref{CIIiv})
if the local monodromy matrices do.
These relations are equivalent to the following
quadratic commutation relations~:
\debut
(u-v)\BBL~T^{ab}(u)~,~T^{cd}(v)~\BBR = \la\(
T^{cb}(u)T^{ad}(v)-T^{cb}(v)T^{ad}(u)\) \label{comrel}
\fin
An important consequence of the relations (\ref{CIIiv})
is that the transfer matrix $\CT(u)$, which is
the trace of the monodromy matrix, $\CT(u)=tr(T(u))= T^{11}(u)+T^{22}(u)$,
is a generating function of commuting hamiltonians~:
\debut
\BBL~ \CT(u)~,~\CT(v)~\BBR=0 \non
\fin
The Heisenberg hamiltonian is recovered
by expanding the logarithm of the trace to first order:
$ H \propto \partial_u \log \CT(u) \Big\vert_{u=0}$.

Another generating function of commuting quantities is given
by the quantum determinant $\det_qT(u)$. It is defined by~:
\debut
\det_qT(u)= T^{22}(u-\la)T^{11}(u)-T^{21}(u-\la)T^{12}(u) \label{qdet}
\fin
It commutes with all the matrix elements of the monodromy matrix~:
$\BBL \det_qT(u),T^{ab}(v)\BBR=0$.

The quadratic algebra (\ref{comrel}) is called a su(2) Yangian \cite{Dr}.
More precisely, consider a $T$-matrix satisfying the commutation
relations (\ref{CIIiv}) or (\ref{comrel}), and normalized
to have a quantum determinant equal to one: $\det_qT(u)=1$.
Assume that the  $T$-matrix possesses a $(\inv{u})$-expansion as
follows~:
\debut
T^{ab}(\lambda) = \de^{ab} + \la \sum_{n=0}^\infty u^{-n-1} t_{(n)}^{ab}
\label{CIIxi}
\fin
Then, the su(2) Yangian is the associative algebra generated
by the elements $t_{(n)}^{ab}$. For these elements, the relations
(\ref{comrel}) are equivalent to~:
\debut
\BBL  t_{(0)}^{ab},t_{(m)}^{cd}  \BBR
&= &\delta^{cb}t_{(m)}^{ad}- \delta^{ad} t_{(m)}^{cb} \label{CIIxiii}\\
\BBL  t_{(n+1)}^{ab},t_{(m)}^{cd}  \BBR
 -\BBL  t_{(n)}^{ab},t_{(m+1)}^{cd}  \BBR
&=& \la\BL  t_{(m)}^{cb}t_{(n)}^{ad}
- t_{(n)}^{cb}t_{(m)}^{ad} \BR \nonumber
\fin
Note that with the quantum determinant constraint,
the $(\inv{u})$-expansion of the
monodromy matrix can be reconstructed from its
two first components $t_{(0)}^{ab}$ and $t_{(1)}^{ab}$.
The relations (\ref{CIIxiii}) clearly shows the Yangians are
deformation of loop algebras.

$\bullet \bullet \bullet$) The next step consists in diagonalizing the
transfer matrix. The algebraic Bethe ansatz provides a way
to perform this diagonalization inside a finite dimensional
irreducible representation of the su(2) Yangian.
Similarly as for the unitary representations of su(2), any finite
dimensional irreducible Yangian representation
is specified by an highest weight vector $\ket{\Om}$.
It is characterized by the following equations~:
\debut
T(u)\ket{\Om}=\pmatrix{ f_1(u) & 0 \cr \star & f_2(u)\cr}\ket{\Om}
\label{hwv}
\fin
where $f_1(u)$ and $f_2(u)$ are C-number functions, not operators.
The product of these functions is related to the quantum determinant
by~: $\det_qT(u)=f_2(u-\la)f_1(u)$.
Due to the fact that the quantum determinant commutes with the
$T$-matrix, only the ratio $f_1(u)/f_2(u)$ encodes the data of
the representation. Moreover, the Yangian representation
with highest weight vector $\ket{\Om}$ is finite dimensional
if and only if this ratio satisfies \cite{Dr}~:
\debut
\frac{f_1(u)}{f_2(u)} = \frac{P(u+\la)}{P(u)} \label{drpol}
\fin
for some polynomial $P(u)$. These polynomials are called
Drinfel'd polynomials. The condition (\ref{drpol}) is
the analogue of the fact that finite dimensional su(2)
representations correspond to half integer spins.

All the states in an irreducible Yangian representation are
obtained by iterative actions of $T^{21}(u)$ on $\ket{\Om}$~:
\debut
\ket{\Psi}= T^{21}(u_1)~T^{21}(u_2)\cdots T^{21}(u_M)\ket{\Om}
\label{Bstate}
\fin
The Bethe states, which are eigenstates of the transfer matrix,
are of this form, but for particular values of the parameters $u_p$.
The relations determining these $u_p$'s are called the
Bethe ansatz equations.
They can be summarized as follows.
Let us define a polynomial $Q(u)$ of degree $M$ whose roots are the $u_p$'s~:
\begin{eqnarray}
Q(u) = \prod_{p=1}^M(u-u_p) \label{qpol}
\end{eqnarray}
The state (\ref{Bstate}) is then  an eigenstate of the transfer matrix
$\CT(u)$ if the roots $u_p$ of $Q(u)$ are such that this polynomial
is solution of the following difference equation~:
\begin{eqnarray}
t(u) Q(u) = f_1(u) Q(u-\la) + f_2(u) Q(u+\la) \label{baxter}
\end{eqnarray}
where $t(u)$, a polynomial of degree $N$, is the eigenvalue
of the transfer matrix on the Bethe state (\ref{Bstate}). Notice that
eq. (\ref{baxter}) at the same time gives the equations
determining the Bethe roots $u_p$ and the eigenvalue $t(u)$.
Eq.(\ref{baxter}) was introduced by Baxter in its solution of
the 8-vertex model \cite{baxt}.

Following an idea due to Sklyanin \cite{Skly}, the
Bethe eigenstates can then be rewritten in terms of the polynomial $Q(u)$.
Since the operator $T^{21}(u)$ is a polynomial of degree $(N-1)$, let us
assume that we can factorized it as follows,
\debut
T^{21}(u) = \la~S^-~.~\prod_{k=1}^{N-1}(u-x_k) \label{Bfact}
\fin
where the $x_k$ are operators and $S^-=\sum_jS_j^{21}$.
It follows from the relations (\ref{comrel}) that
the $x_k$ are commuting operators.
The Bethe eigenstates (\ref{Bstate}) are then given by
\debut
\ket{\Psi} = \(S^-\)^M~ Q(x_1)~Q(x_2)~\cdots~Q(x_{N-1})\ket{\Om}
\label{bethe}
\fin
The eqs.(\ref{baxter},\ref{bethe}) reflect the separation of the variables,
since the eigenstates are determined from the solutions of one
equation for a function of one variable only.

Finalizing the solution of the models consists in
analyzing the Bethe ansatz equations and their solutions. This can
analytically be done only in the thermodynamical limit, along
the lines outlined in section 1.3.

\subsection{Quantization of the spectral parameter.}

The long-range interacting models cannot be solved using the algebraic
Bethe ansatz. This follows from the fact the hamiltonian commutes with
the $T$-matrix, and therefore non-degenerate eigenstates cannot be obtained by
iterative action of the lowering operators $T^{21}(u)$.
Nevertheless, the tools of the algebraic Bethe ansatz are useful for
constructing integrable long range interacting models and for
deciphering the symmetries of these models.

To illustrate this fact, we now consider su(2) generalizations of the
Calogero-Sutherland models.
These models describe $M$ particles interacting by long range forces.
Their positions are parameterized by complex numbers $z_i$,
$i=1,\cdots,M$, and each particle carries a spin $\sigma=\pm$.
The Hamiltonian is~:
\eqn{ H_D = \sum_{j=1}^M (z_j\d_{z_j})^2 - \sum_{i\not= j}
        \la(P_{ij}+\la)  \frac { z_i z_j}{ (z_i-z_j)^2}
\label{Iiii} }
where $\la$ is a coupling constant and $P_{ij}$ exchanges the spins
of the particles $i$ and $j$. Notice we recover the Haldane-Shastry
spin chain in the static limit $\la=\infty$.

The construction of these models relies on the definition
a monodromy matrix in which the spectral parameter has been quantized.
More precisely, let us consider the monodromy matrix (\ref{CIIIiii})
but in which the spectral parameters have been shifted to $(u-\hat D_i)$,
where the $\hat D_i$ are operators, commuting among themselves and
with the spin operators. More precisely,
we define a $\hat T$-matrix by \cite{long}~:
\eqn{
\hat T^{ab}(u) = \sum_{a_2\cdots a_N}\
\hat T_1^{a a_2}(u)~\hat T_2^{a_2 a_3}(u)~\cdots~ \hat T_N^{a_N b}(u)
\label{That} }
with
\eqn{ T^{ab}_i(u) = \frac{ (u-\hat D_i)\de^{ab}+\la S_i^{ab}}{u-\hat D_i}
\label{Wiii} }
The operators $\hat D_i$ we consider are defined as follows \cite{long}~:
\eqn{
\hat D_i = z_i\d_{z_i} + \la \sum_{j>i} \theta_{ij} K_{ij}
- \la \sum_{j<i} \theta_{ji} K_{ij}
\label{Wi} }
where $\th_{ij}=\frac{z_i}{z_i-z_j}$ and $K_{ij}$ the operators
which exchange the particles at positions $z_i$ and $z_j$~:
$K_{ij}z_j=z_iK_{ij}$.
They obey the defining relations of a degenerate affine Hecke algebra~:
\eqn{
\BBL\ \hat D_i\ ,\ \hat D_j\ \BBR\ &=& 0 \non \\
\BBL\  K_{i,i+1}\ ,\hat D_k\ \BBR\ &=&\ 0 \quad {\rm if\ } k \not= i,i+1 \non\\
K_{i,i+1} \hat D_i - \hat D_{i+1} K_{i,i+1} &=&\ -\la
\label{hecke} }
In the mathematics literature, the role of the affine Hecke algebra
in this context was revealed by Cherednik \cite{Chred}.
In the physics literature, operators similar but different to the $\hat D_i$
were introduced by Polykronakos \cite{Poly}.
Notice that these relations imply that~:
\eqn{
\[ K_{ij},\hat \Delta_M(u)\]=0,\quad {\rm with}\quad
\hat \Delta_M(u)=\prod_{i=1}^M (u-\hat D_i)
\label{Wv} }
I.e. $\hat \Delta_M(u)$ is symmetric by permutation of the particles.
This property follows from
$\[ K_{ii+1},(u-\hat D_i)(u-\hat D_{i+1})\] = 0$, which is
valid for all $i$.

Since the operators $\hat D_i$ commute,
the $\hat T$-matrix (\ref{That}) satisfies the $RTT$ relation (\ref{CIIiv}).
However, the positions and the spin variables are totally uncoupled
since the operators $\hat D_i$ commute with the spin operators.
In order to couple them, we define a projection $\pi$ which consists
in replacing the permutation $K_{ij}$ by the permutation $P_{ij}$
after it has been moved to the right of an expression.
One can view this projection as the result of acting on
wave functions totally symmetric under simultaneous permutations
of the positions and of the spins. In more mathematical words, this
procedure consists in quotienting the algebra generated by the
permutations $K_{ij}$ and $P_{ij}$ by the left ideal generated by
$(K_{ij}-P_{ij})$.  We use it to eliminate the permutations of the particles
by replacing them with those of the spins.

The transfer matrix $T(u)$ defined by
\eqn{
T(u)=\pi(\hat T(u)) \label{piT} }
will then satisfy the Yang-Baxter equation
if we can replace the projection of the product
$(1\otimes \hat T(v))(\hat T(u)\otimes 1)$ by the product of the projections.
Since, $\hat \De_M(u)$ is symmetric under permutation, it is equivalent
to check this property for $\hat T'(u)=\hat \De_M(u)\hat T(u)$.
For this to be true, $\hat T(u) $ applied on a totally symmetric
wave function must still be a totally symmetric wave function. Equivalently,
we must have:
\eqn{  \pi\({ K_{ij} \hat T(u) }\) = P_{ij} \pi \({ \hat T(u) }\)
\label{fond} }
Since the permutation groups are generated by the permutations $K_{i,i+1}$
and $P_{i,i+1}$, eq.(\ref{fond}) is equivalent to~:
$\pi \( K_{i,i+1} \hat T_i(u) \hat T_{i+1}(u)\) =
P_{i,i+1} \pi\(\hat T_i(u)\hat T_{i+1}(u)\)$, with
$\hat T_i(u)$ defined in (\ref{Wiii}).
This is garanteed if the commutation relations of the degenerate
Hecke algebra (\ref{hecke}) are satisfied. Thus, the relations (\ref{hecke})
are
the necessary relations for this $T$-matrix to satisfy the $RTT$-relation.

An alternative presentation of this $T$-matrix
was obtained in ref.\cite{long}~:
\eqn{
T^{ab}(u)\ =\  \de^{ab} + \la \sum_{i,j=1}^M S_i^{ab}\({\inv{u-L}}\)_{ij}
\label{IViii} }
where $L$ is the matrix defined by~:
$L_{ij} = \de_{ij} z_j\d_{z_j} + (1-\de_{ij})\la \th_{ij}P_{ij}$,
with $\th_{ij}=z_i/(z_i-z_j)$.
In eq.(\ref{IViii}), the projection $\pi$ has been explicitely done.

The immediate consequences of this construction are the following.
Since the $T$-matrix (\ref{piT}) satisfies the relation (\ref{CIIiv})
it defines a representation of the su(2) Yangian. As explained in
the previous section, the relation (\ref{CIIiv}) implies that
$\CT(u)=tr(T(u))$ is a generating function of commuting hamiltonian.
However, $\CT(u)$ is not Yangian invariant since it does not commute
with $T$ itself. A clever choice consists in choising the quantum
determinant $\det_qT(u)$ as the generating function of commuting hamiltonians.
It is the projection of the quantum determinant of $\hat T(u)$:
\eqn{
\det_q T(u)=\ \pi\(\frac{\hat \De_M(u+\la)}{\hat \De_M(u)}\)
\label{Wiv} }
where $\hat \Delta_M(u)$ is defined in eq.(\ref{Wv}).
The hamiltonian (\ref{Iiii}) is the $u^{-2}$-term in (\ref{Wiv}). It is
therefore Yangian invariant. This invariance has recently been
imbeded in a bigger algebra which is a deformation a $W_\infty$ algebra
\cite{wada}.
The quantum determinant (\ref{Wiv}) has been diagonalized in ref.\cite{long}
by directly diagonalizing the operators $\hat D_i$.

\subsection{Application to the Haldane-Shastry spin chain.}

We now explain how the previous construction can be used to derive
the fractional selection rules satisfied by the eigenstates of the
Haldane-Shastry spin chain.

As mentioned in section 2, the  Haldane-Shastry spin chain is
Yangian invariant. Therefore, there exists a $T$-matrix
commuting with the hamiltonian (\ref{EAa}) and
satisfying the relations (\ref{CIIiv}).
It was constructed in \cite{long}. It is the
limit $u,\la\to\infty$ with $x=u/\la$ fixed, of the $T$-matrix (\ref{IViii}).
Its expression is~:
\eqn{
 T^{ab}(x)= \de^{ab} +\sum_{i,j=1}^N S_i^{ab}\Bigl({\inv{x-L'} }\Bigr)_{ij}
\label{EBb} }
with $L'_{ij}=(1-\de_{ij})\th_{ij}P_{ij}$,
$\theta_{ij}=z_i/z_{ij}$ with $z_{ij}=z_i-z_j$, and $S_i^{ab}$ is the canonical
matrix $\ket{a}\bra{b}$ acting on the $i^{th}$ spin only.
For any values of the complex numbers $z_j$,
the transfer matrix (\ref{EBb}) form a representation of the exchange
algebra (\ref{comrel}) with $u$ changed into $x$ and $\la$ normalized
to one.  The trigonometric spin chain corresponds to $z_j=\om^j$ with
$\om$ a primitive $N^{th}$ root of the unity.  For these values
of $z_j$, the transfer matrix (\ref{EBb}) commutes with
the hamiltonian (\ref{EAa}).

In the representation (\ref{EBb}), the quantum determinant is a pure number
for any values of the $z_j$'s given by~:
\eqn{
 det_q\ T(x)\ =\ 1 + \sum_{i,j=1}^N \Bigl({ \inv{x-\Theta}}\Bigr)_{ij}
 =\ \frac{ \De_N(x+1) }{\De_N(x) }
\label{EBd} }
with $\De_N(x) $ the characteristic polynomial of the
$N\times N$ matrix $\Theta$ with entries $\th_{ij}$:
$\De_{N}(x)= {\rm det}(x-\Theta)$. For the Haldane-Shastry spin chain
$z_j=\om^j$ and we have~:
\eqn{
\De_N(x)= \prod_{j=1}^N \bigl( x+\frac{N+1}{2} -j \bigr)
\label{EBz} }

Since the monodromy matrix (\ref{EBb}) commutes with the hamiltonian,
the long-range interacting spin chain cannot be solved using the algebraic
Bethe ansatz. A way to solve it consists first in decomposing the
spin chain Hilbert space into irreducible sub-representation of
the Yangian, and then in computing the energy in each of these
irreducible blocks. For the values
$z_j=\om^j$, the representation (\ref{EBb}) is completely reducible.
Each irreducible sub-representation
possesses a unique highest weight vector $\ket{\La}$ which
is annihilated by $T^{12}(x)$ and which is an eigenvector of the
diagonal components of the transfer matrix, as in eq.(\ref{hwv}).
In ref.\cite{long}, it was shown that the corresponding eigenvalues
of $T^{11}(x)$ and $T^{22}(x)$ can be expressed in terms of two
polynomials $P(x)$ and $Q(x)$~:
\eqn{
T(x)\ket{\La} = \frac{Q(x+1)}{Q(x)}
        \pmatrix{ \frac{P(x+1)}{P(x)} & 0\cr
                                ~ & ~ \cr
                        \star & 1 \cr} \ket{\La}
\label{EDk} }
These polynomials characterize the irreducible sub-representations.
The polynomials $Q(x)$ and $P(x)$ are not independent,
since the quantum determinant (\ref{EBb}) take the same value in any of the
irreducible blocks. They should satisfy~:
\eqn{\Delta_N(x) = P(x)\ Q(x)Q(x-1).
\label{EDj} }
Therefore, the roots of $P(x)$ and $Q(x)$ are among those of $\Delta_N(x)$.
This implies that $Q(x)$ factorized as~:
\eqn{
Q(x) = \prod_{p=1}^M(x+\frac{N+1}{2}-k_p)
\label{EDi} }
where the $\{k_p\}$ are integers bewteen $1$ and $(N-1)$.
The equation (\ref{EDj}) then admits solutions
if and only if the roots of $Q(x)$ are not adjacent, or equivalently,
if and only if the integers $\{k_p\}$ are neither equal nor adjacent.
These integers will be identified with the rapidities labeling
the eigenmultiplets of the spin chain.

This provides a purely algebraic way to recover the rapidity selection rule.
It also shows that the fractional statistics of the spinon excitations
is an echo of the Yangian symmetry.

\bigskip \bigskip
\noindent {\bf Acknowledgements:} It is a pleasure to thank
F. David and P. Ginsparg for having organized this school.
I also would like to thank my collaborators,
O. Babelon, D. Haldane, V. Pasquier,  D. Serban and Y.-S. Wu,
for having shared their enthusiasm on this subject.

\end{document}